\documentclass[final]{svjour2}
\usepackage{graphicx}
\usepackage{rotating}
\usepackage{amsmath}
\usepackage[numbers]{natbib}
\makeatletter
\journalname{Journal of Low Temperature Physics}

\bibpunct{}{}{,}{s}{}{,}
\usepackage{bm}
\begin{document}

\title{Hydrodynamic Instability and Turbulence in Quantum Fluids}
\author{Makoto Tsubota}


\institute{M. Tsubota 
               \at Department of Physics and the OCU Advanced Research Institute for Natural Science and Technology (OCARINA), Osaka City University, Sumiyoshi-ku, Osaka 558-8585, Japan \\
               \email{tsubota@sci.osaka-cu.ac.jp} \\ 
			   \\}

\date{Received: date / Accepted: date}

\maketitle


\begin{abstract}
Superfluid turbulence consisting of quantized vortices is called quantum turbulence (QT). 
Quantum turbulence and quantized vortices were discovered in superfluid $^4$He about 50 years ago,
but innovation has occurred recently in this field. 
One is in the field of superfluid helium.  
Statistical quantities such as energy spectra and probability distribution function of the velocity field have been
accessible both experimentally and numerically.
Visualization technique has developed and succeeded in the direct visualization of quantized vortices.
The other innovation is in the field of atomic Bose-Einstein condensation. 
The modern optical technique has enabled us to control and visualize directly the condensate and quantized vortices. 
Various kinds of hydrodynamic instability have been revealed. 
Even QT is realized experimentally. 
This article describes such recent developments as well as the motivation of studying QT.


\end{abstract}

\section{Introduction}
Quantum turbulence (QT) refers generally to turbulence consisting of quantized vortices of superfluid component. 
Quantum turbulence and quantized vortices were discovered in superfluid $^4$He about 50 years ago, 
and have been thoroughly studied experimentally, theoretically and numerically. 
Since the early efforts were mostly limited to thermal counterflow, however, 
QT has absorbed little interest of scientists in other fields.
Innovation has occurred in the research field, namely one in superfluid helium and the other in atomic Bose-Einstein condensates (BECs).

In this article we review the innovation following the basic motivation.
Section 2 describes the basic motivation of this field.
Sections 3 and 4 discuss the recent developments in superfluid helium and atomic BECs respectively.
Section 5 is devoted to conclusions.
The topics in this field are too diverse to be covered by this article. 
The readers should refer to some other review articles for the whole story
\cite{Hal,TsubotaJPSJ,Vinen10,Skrbek2012,TK12}.

 \section{Motivation of studying quantum hydrodynamics and turbulence}
Quantum hydrodynamics has been studied in connection with superfluid helium for over 60 years.
Liquid $^4$He enters a superfluid state below the $\lambda$ point (2.17 K) through
Bose--Einstein condensation of the atoms. 
The characteristics of superfluidity were discovered experimentally
in the 1930s by Kapitza  et al.
The hydrodynamics of superfluid helium is well described by the two-fluid model,
in which the system consists of an inviscid superfluid of density $\rho_s$ and 
a viscous normal fluid of density $\rho_n$ with two independent velocity fields $\mathbf{v}_s$ and $\mathbf{v}_n$. 
The mixing ratio of the two fluids depends on temperature. 
As the system is cooled down below the $\lambda$ point, the ratio of 
the superfluid component increases, and the entire fluid becomes
superfluid below about 1 K.
The Bose-condensed system exhibits the macroscopic wave function
$\Psi(\mathbf{x},t)=|\Psi(\mathbf{x},t)| e^{i \theta(\mathbf{x},t)}$ as an order parameter. 
The superfluid velocity field representing the potential flow is given by $\mathbf{v}_s=(\hbar/m) \nabla \theta$ with boson mass $m$. 
Since the macroscopic wave function must be single-valued for space coordinate $\mathbf{x}$, the circulation $\Gamma = \oint \mathbf{v} \cdot d\mathbf{\ell}$ for an arbitrary closed loop within the fluid is quantized in terms of the quantum value $\kappa=h/m$.  A vortex with such quantized circulation is called a quantized vortex. Any rotational motion of a superfluid is only sustained 
by quantized vortices.

A quantized vortex is a topological defect characteristic of a Bose--Einstein condensate and is different from a vortex in a classical viscous fluid. First, its circulation is quantized and conserved, in contrast to a classical vortex whose circulation can have any value and is not conserved. Second, a quantized vortex is a vortex of inviscid superflow. Thus, it cannot decay by the viscous diffusion of vorticity that occurs in a classical fluid.  Third, the core of a quantized vortex is very thin, on the order of the coherence length, which is only a few angstroms in superfluid $^4$He and submicron in atomic BECs. Since the vortex core is thin and does not decay by diffusion, it is possible to identify the position of a quantized vortex in the fluid. These properties make a quantized vortex more stable and definite than a classical vortex. 

Early experimental studies of superfluid turbulence focused primarily on thermal counterflow of superfluid $^4$He, in which the normal fluid and superfluid flow in opposite directions. The flow is driven by an injected heat current, and it is found that the superflow becomes dissipative when the relative velocity between the two fluids exceeds a critical value \cite{Gorter}. Feynman proposed that it is a superfluid turbulent state consisting of a tangle of quantized vortices \cite{Feynman}. Vinen later confirmed Feynman's proposal experimentally by showing that the dissipation arises from mutual friction between vortices and the normal flow \cite{Vinen1957a, Vinen1957b, Vinen1957c, Vinen1957d}. The quantization of circulation was also observed by Vinen using a vibrating wire technique \cite{Vinen1961}. Subsequently, many experimental studies have examined superfluid turbulence (ST) in thermal counterflow systems, revealing a variety of physical phenomena \cite{Tough1982}. The dynamics of quantized vortices are nonlinear and nonlocal, so it has not been easy to quantitatively understand the experimental results on the basis of vortex dynamics. A breakthrough was achieved by Schwarz, who clarified the picture of ST consisting of tangled vortices by a numerical simulation of the quantized vortex filament model in the thermal counterflow \cite{Schwarz1985, Schwarz1988}. However, since the thermal counterflow has no analogy with conventional fluid dynamics, these experimental and numerical studies are not helpful in clarifying the relationship between ST and classical turbulence (CT). Superfluid turbulence is often called quantum turbulence (QT), which emphasizes the fact that it is comprised of quantized vortices.

Comparing QT and CT reveals definite differences. Turbulence in a classical viscous fluid appears to be comprised of vortices. However, these vortices are unstable and not well defined. Moreover, the circulation is not conserved and is not identical for each vortex. Quantum turbulence consists of a tangle of quantized vortices that have the same conserved circulation.  Looking back at the history of science, {\it reductionism}, which tries to understand the nature of complex things by reducing them to the interactions of their parts, has played an extremely important role. The success of solid-state physics owes much to  {\it reductionism}. In contrast, conventional fluid physics is not reducible to elements, and thus does not enjoy the benefits of {\it reductionism}. However, quantum turbulence is different, being reduced to quantized vortices. Thus {\it reductionism} is applicable to quantum turbulence.  

Two formulations are generally available for studying the dynamics of quantized vortices.  
One is the vortex filament model and the other is the Gross--Pitaevskii (GP) model.
The vortex filament model represents a quantized vortex as a filament passing through the fluid, having a definite direction corresponding to its vorticity. 
Except for the thin core region, the superflow velocity field has a classically well-defined meaning and can be described by ideal fluid dynamics. 
The velocity at a point due to a filament is given by the Biot--Savart expression. 
The GP model describes a weakly interacting Bose system. The macroscopic wave function $\Psi(\mathbf{x},t)$ appears as the order parameter of Bose--Einstein condensation, obeying the Gross--Pitaevskii (GP) equation \cite{Pethick}: 
\begin{equation}
i \hbar \frac{\partial \,\Psi(\mathbf{x},t)}{\partial \,t} = \biggl( - \frac{\hbar ^2}{2m}\nabla^2
+g |\Psi(\mathbf{x},t)|^{2}-\mu \biggr) \Psi(\mathbf{x},t),
\label{gpeq}
\end{equation}
where $g$ is the interaction parameter and $\mu$ is the chemical potential.
 Here it should be noted that the GP model is not quantitatively applicable to superfluid $^4$He that is a strongly correlated system.
 
\section{Superfluid helium}
Since the middle of 1990's two kinds of important developments have occurred in the field of superfluid helium. One is classical analogue, and the other is visualization of quantized vortices and QT. The recent developments in the numerical simulation of the vortex dynamics are also described.

\subsection{Classical analogue}
Since thermal counterflow has no classical analogue, understanding the correspondence between QT and CT has been deserted. 

One of the key issues for studying turbulence is to focus on statistical quantities \cite{Frisch}.
The most important statistical law is the Kolmogorov law \cite{K41}, which states that the spectrum of the kinetic energy exhibits the $-5/3$ power law in the fully developed isotropic homogeneous turbulence.. 
Lots of experiments for the situations apart from counterflow have been performed in superfluid $^4$He and the B-phase of superfluid $^3$He to obtain the results consistent with the energy spectra with the Kolmogorov-like power law. The readers should refer to other review articles about these experiments \cite{Hal, Vinen10}.

Here we briefly review some numerical and theoretical works on these topics. 
We confine ourselves to the system at 0K.
The current understanding is the follows.
The energy spectrum and the vortex cascade process are usually divided into two regions in wavenumber space.

The first region is called the semiclassical region, existing at wavenumbers below the inverse of the mean intervortex spacing.
The dynamics of vortices in the classical region are dominated by the Richardson cascade, in which large vortices are broken up self-similarly into smaller ones, or collective dynamics of aggregated quantized vortices at scales larger than the intervortex spacing.
Such behavior of vortices supports the analogy of QT to CT, namely, the Kolmogorov energy spectrum, which has been confirmed by several numerical efforts of both the vortex filament model and the GP model \cite{Hal}.
However, most of these works dealt with the relatively narrow range of wave numbers, typically only one order, chiefly because of the numerical constraint.
 The recent large simulation of the GP model treats the wider range of the wave number \cite{Sasa2011}, observing that the $-5/3$ power law is followed by an energy accumulation at about the reciprocal mean intervortex spacing. 
This behavior seems consistent with the bottleneck model \cite{LNR2008}.   
 
The second region is called the quantum region at wavenumbers above the inverse of the mean intervortex spacing, in which vortex dynamics are dominated by the effects of the quantized circulation, specifically the Kelvin wave cascade of vortices, which does not appear in CT.
The Kelvin wave cascade is also a very important concept in understanding the dissipation mechanism of QT at very low temperatures.

The above is the current theoretical understanding, but we have no definite experimental proof.

\subsection{Visualization of quantized vortices and QT}
There has been little direct experimental information about the flow in superfluid $^4$He. This is mainly because usual flow visualization techniques are not applicable to superfluid. However, the situation changes rapidly \cite{SciverPLTP}. For QT, we should seed the fluid with tracer particles in order to visualize the flow field and, if possible, quantized vortices, which are observable by appropriate optical techniques.  

Three kinds of tracers are currently used for the visualization.
 Using the particle image velocimetry (PIV) technique with 1.7-$\mu$m-diameter polymer particles, Zhang and Van Sciver visualized a large-scale turbulent flow both in front of and behind a cylinder in a counterflow in superfluid $^4$He at finite temperatures \cite{Zhang05}.  
 Bewley et al. seeded \cite{Bewley06}  liquid helium with solid hydrogen particles smaller than 2.7 $\mu$m.  When the temperature is above $T_\lambda$, the particles were seen to form a homogeneous cloud that disperses throughout the fluid. However, on passing through $T_\lambda$, the particles coalesced into web-like structures, probably because they were trapped by quantized vortices.   
Then Paoletti et al. observed vortex reconnections and some characteristic scaling law \cite{Paoletti2010}, and visualized thermal counterflow \cite{Paoletti2008}. 
Using metastable helium molecules,  Guo et al. visualized the normal flow in thermal counterflow, thus finding that the normal flow is also turbulent at relatively large velocities
 \cite{Guo2010}. 

\subsection{Numerical simulation of the vortex filament model by the full Biot-Savart law}

Numerical simulation of the vortex filament model for superfluid $^4$He was developed by Schwarz \cite{Schwarz1985}, who succeeded in obtaining a statistically steady state in thermal couterflow under periodic boundary conditions \cite{Schwarz1988}.
However,  the numerical simulation had serious defects.
 One is that the calculations were performed under the localized induction approximation (LIA) neglecting interactions between vortices in the Biot-Savart law. 
 Schwarz reported that as a result the layer structure was constructed gradually. 
 Of course, this behavior is not realistic.
 In order to address this, an unphysical, artificial mixing procedure was employed, in which half the vortices are randomly selected to be rotated by 90$^\circ$ around the axis defined by the flow velocity. 
 
\begin{figure}[htb] \centering \begin{minipage}{0.49\linewidth} \begin{center} \includegraphics[width=.99\linewidth]{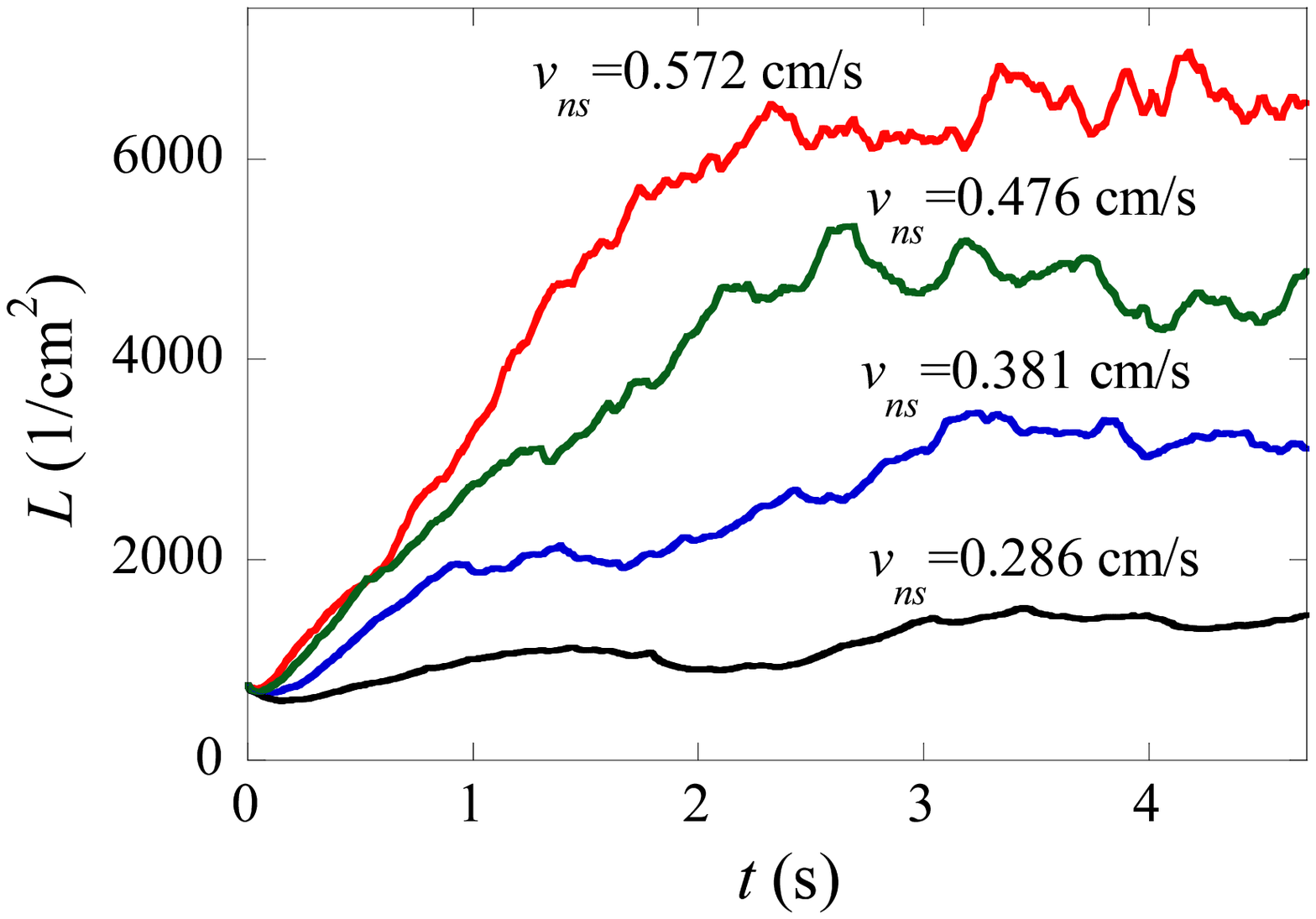}\\ (a) \end{center} \end{minipage} \begin{minipage}{0.49\linewidth} \begin{center} \includegraphics[width=.99\linewidth]{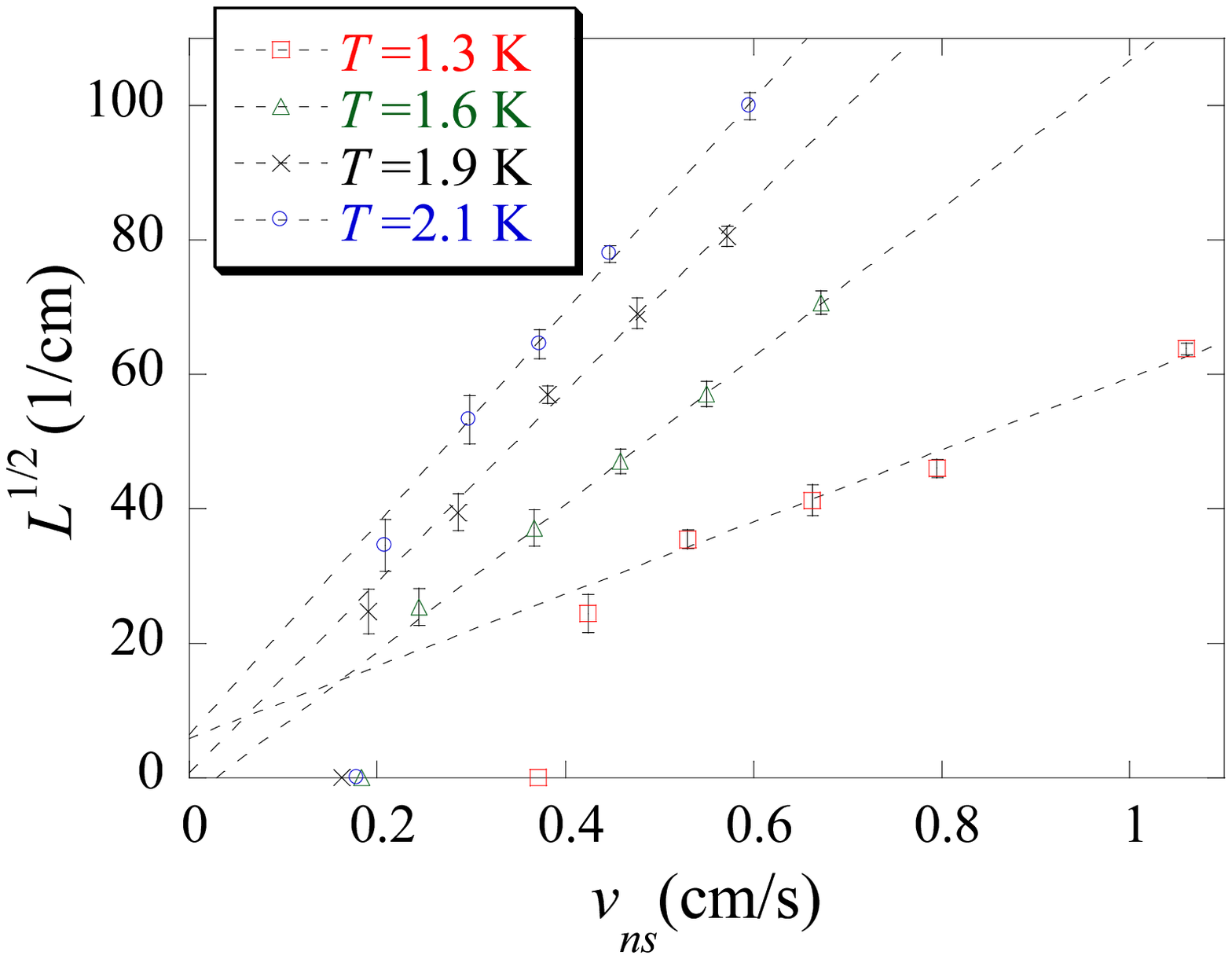}\\ (b) \end{center} \end{minipage}
\caption{(Color online) Numerical results for the simulation of the vortex filament model for a computing box of 0.1 $\times$ 0.1 $\times$ 0.1 cm$^3$. (a) Vortex line density as a function of time for four different counterflow velocities.  (b) Steady state vortex line density $L(t)$ as a function of the counterflow velocity $v_{ns}$. The error bars represent the standard deviation. [Adachi, Fujiyama and Tsubota: Phys. Rev. B \textbf{81} (2010) 104511, reproduced with permission. Copyright 2010 the American Physical Society.]  } \label{Adachi} \end{figure}

 Adachi et al. performed numerical simulations of counterflow turbulence using the full Biot--Savart law under periodic boundary conditions and succeeded in obtaining a statistically steady state without any unphysical procedures \cite{Adachi2010}.
Starting with six vortex rings, they followed the vortex dynamics under the counterflow velocity $v_{ns}$.
Differently from the LIA simulation, the intervortex interaction prevents the vortices from forming the layer structure, thus realizing a statistically steady state by the competition between the growth and decay of a vortex tangle through mutual friction.
Figure \ref{Adachi} (a) shows how the  vortex line density (VLD) $L$, namely the total length of vortex lines per unit volume, grows towards a statistically steady state.
The obtained steady states almost satisfy the relation of $L=\gamma^2v_{ns}^2$ when $v_{ns}$ and $L$ are relatively large, as shown in Fig. \ref{Adachi} (b).
This relation is confirmed by a large number of the observations of stationary cases \cite{Tough1982}.
The temperature-dependent parameter $\gamma$ is obtained numerically, agreeing quantitatively with some typical observations.
Thus a full account of the intervortex interaction by the full Biot--Savart law enables us to obtain the statistical steady states without any unphysical procedures.
A detailed comparison between the numerical results of the full Biot--Savart law and the LIA is discussed in Ref. 27.

After the work, the numerical works of the vortex filament model based on the full Biot--Savart law make progress. 
For example, reconnections of vortex bundles are studied by the full Biot-Savart filament model as well as the GP model \cite{Alamri2008}.  

\section{Atomic Bose-Einstein condensates}
The realization of Bose--Einstein condensation in trapped atomic gases in 1995 provided another important system for studying quantum hydrodynamics \cite{Fetter2009,KasaPLTP}.
The existence of superfluidity has been confirmed by creating and observing quantized vortices in atomic BECs. 
Atomic BECs have several advantages over superfluid helium,
the most important being that modern optical techniques can be used to directly control their properties and to visualize quantized vortices.
However, most efforts have been devoted to small number of vortices or vortex lattice under rotation.  
This section reviews briefly the recent progresses on hydrodynamics instability and QT in atomic BECs.

\subsection{Creating QT}
It is not so trivial how to make QT in a trapped BEC.
A few methods have been proposed theoretically and numerically.
Berloff and Svistunov showed that uniform BECs started from a strongly nonequilibrium state to develop to QT \cite{Berloff2002}.
Parker and Adams suggested the emergence and decay of turbulence in a trapped BEC under a simple rotation, starting from a vortex-free equilibrium BEC \cite{Parker2005}.
A numerical simulation based on the GP equation suggests the appearance of a turbulent regime that contains many vortices and density fluctuations (sound waves) on a route to the ordered vortex lattice.

\begin{figure}
\begin{center}
\includegraphics[%
  width=0.80\linewidth,
  keepaspectratio]{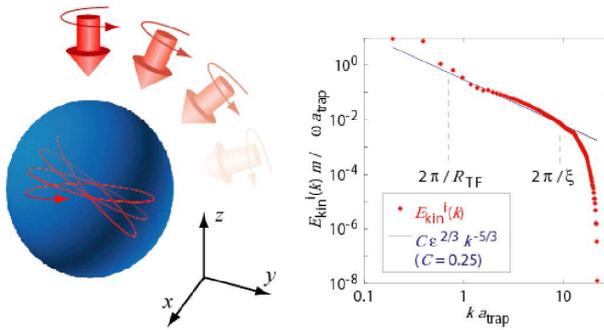}
\end{center}
\caption{(Color online) QT in atomic BECs. 
(a) A method for realizing QT in an atomic BEC subject to precession. 
A BEC is trapped in a weakly elliptical harmonic potential. 
A rotation is applied along the $z$-axis followed by a rotation along the $x$-axis. 
(b) Energy spectrum of steady QT obtained by two-axis rotation. 
The dots represent the numerically obtained spectrum for an incompressible kinetic energy, 
while the solid line is the Kolmogorov spectrum. Here, $R_{\rm TF}$ is the 
Thomas--Fermi radius, $a_{\rm trap}=\sqrt{\hbar/m\omega}$ is the characteristic length scale of the trap and $\xi$ is the coherence length.  
[Kobayashi and Tsubota: Phys. Rev. A {\bf 76}, 045603 (2007), reproduced with permission.
Copyright 2007 by the American Physical Society].}
\label{Kobayashi}
\end{figure}

Kobayashi and Tsubota suggested rotations about two axes \cite{Kobayashi2007}, 
as shown in Fig. \ref{Kobayashi}(a). 
The simulation was performed by the GP model with a dissipative term.
The numerically obtained spectrum for an incompressible kinetic energy showed a Kolmogorov-like power law (Fig. \ref{Kobayashi}(b)).

Being motivated by the theoretical work, Henn et al. introduced an external oscillatory potential to an trapped $^{87}$Rb BEC and succeeded in making a vortex tangle \cite{Henn2009a,Henn2009b}.
This oscillatory field induced a successive coherent mode excitation in a BEC.
They observed that increasing the amplitude of the oscillating field and the excitation period increased the number of vortices and eventually lead to the turbulent state \cite{Henn2009b}.
They found also the characteristic free expansion of turbulent clouds after turning off the trapping potential.
In contrast to the isotropic expansion of a thermal cloud, an initially anisotropic BEC cloud expands faster in the most confined directions than in the less confined, so a cigar-shaped cloud expands to turn into a pancake-shaped cloud with an inversion of its aspect ratio.
However, turbulent BEC clouds were observed to expand with keeping the aspect ratio. 
This behavior looks to be a signature of QT, but being not yet understood well. 

\subsection{Dynamics of vortices}
Dynamics of a small number of vortices are investigated.
Neely et al. nucleated vortex dipoles in an oblate BEC by forcing superfluid flow around a repulsive Gaussian obstacle generated by a focused blue-detuned laser beam \cite{Neely}.
The nucleated vortex dipole propagates in a BEC cloud for many seconds; its continuous trajectory was found to be consistent with a numerical simulation of the GP model.
Freilich  et al. observed the real-time dynamics of vortices by repeatedly extracting, expanding, and imaging small fractions of the condensate to visualize the motion of the vortex cores \cite{Freilich}. 
They nucleated vortices via the Kibble--Zurek mechanism in which rapid quenching of a cold thermal gas through the BEC phase transition causes topological defects to nucleate \cite{Weiler}. 

Successive generation of vortices is studied too.
Sasaki et al. studied vortex shedding from an obstacle potential moving in a uniform BEC \cite{Sasaki10}. 
The flow around the obstacle is laminar when the velocity of the potential is sufficiently low.
When the velocity exceeds a critical velocity of the order of the sound velocity, the potential commences to emit vortex trains.
The manner in which vortices are emitted depends on the velocity and the width of the potential.
Aioi et al. subsequently proposed controlled generation and manipulation of vortex dipoles by using several 
Gaussian beams from a red (attractive potential) or blue (repulsive potential) detuned laser \cite{Aioi11}.

In the field of superfluid $^4$He and $^3$He, several groups have experimentally studied QT generated by oscillating structures such as wires, spheres, and grids \cite{SVPLTP}.
This strategy can also be applied to trapped atomic BECs.
Fujimoto and Tsubota numerically investigated the two-dimensional dynamics of trapped BECs 
induced by an oscillating repulsive Gaussian potential \cite{Fujimoto2010, Fujimoto2011} and
found a strong dependence on the amplitude and frequency of the potential.
Unlike a potential with constant velocity \cite{Sasaki10}, an oscillating potential continually 
sheds vortex pairs with alternating impulses.
The nucleated pairs form new vortex pairs through reconnections that move away 
from the obstacle potential and toward the condensate surface.
The BEC cloud eventually becomes full with such vortices.


\subsection{Multi-component BECs}
Multicomponent atomic BECs can be created in cold-atom systems with, for example, multiple hyperfine spin states or a mixture of different atomic species. 
Such systems yield a rich variety of superfluid dynamics.

Two-component BECs are characterized by the intracomponent interaction parameters $g_{11}$ and $g_{22}$ and the intercomponent interaction parameter $g_{12}$ \cite{Kasamatsu05}.  
When $g_{11}g_{22}>g_{12}^2$ and $g_{11}g_{22}<g_{12}^2$, respectively, two condensates are miscible and immiscible, leading to different kinds of hydrodynamic instability.

\begin{figure}
\begin{center}
\includegraphics[%
  width=0.80\linewidth,
  keepaspectratio]{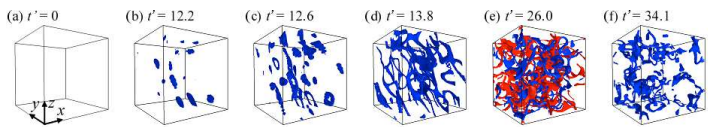}
\end{center}
\caption{(Color online) Characteristic nonlinear dynamics 
of CSI of two-component BECs.
The parameters are set to $m_1=m_2=m$, 
$g_{11}=g_{22}=g$, and $g_{12}=0.9g$. The initial state 
is prepared by adding small random noise to the stationary wave function 
under ${\bm V}_1=-{\bm V}_2$.
The panels show the time development of low-density isosurfaces.
Because of the symmetric parameter setting, the second component 
behaves in a similar manner to the first component (not shown)  
[Takeuchi {\it et al.}: Phys. Rev. Lett. {\bf 105}, 205301 (2010), reproduced with permission.
Copyright 2010 by the American Physical Society].}
\label{CSIfig}
\end{figure}

When two BECs are miscible and flow with different velocities ${\bm V}_1$ and ${\bm V}_2$ , counter-superflow instability (CSI) can occur above a critical relative velocity \cite{Law}. 
Takeuchi et al. suggested that the nonlinear dynamics triggered by the CSI generates a binary QT \cite{TakeuchiCSI,Ishino}. 
Figure \ref{CSIfig} depicts the characteristic nonlinear dynamics of the CSI in a uniform two-component BEC obtained 
by numerically solving the coupled GP equations for two BECs.
In the early stage of the dynamics, amplification of the unstable modes creates disk-shaped low-density regions
that are orientated in the $x$ direction (counterflow direction) [Fig. \ref{CSIfig}(b)].
Each low-density region creates a local dark soliton.
The soliton in the $i$th component transforms into a vortex ring via snake instability  
with a momentum antiparallel to the initial velocity ${\bm V}_i$ [Fig. \ref{CSIfig}(c)].
The vortex rings grow and expand through reconnections, resulting in binary QT in which the vortices of both components are tangled with each other [Fig. \ref{CSIfig}(f)].
The most characteristic is the momentum exchange between two BECs.
At first, two BECs flow as independent superfluids. 
However, as the vortices are created and grow, two BECs start to exchange momentum through vortices, thus reducing their relative motion. 
Such relative motion of two BECs and the initial instability are experimentally realized \cite{Hamner}, but binary QT is not yet observed.

When two BECs are phase-separated, we can expect some kinds of characteristic instability.
One of them is Kelvin-Helmholtz instability (KHI) \cite{TakeuchiKH}, which refers to the surface instability with the shear flow of two BECs.
When the relative velocity exceeds some critical value, the system becomes dynamically unstable against excitation of some interface modes.
This story is similar to classical KHI, but quantum KHI emits quantized vortices from the amplified unstable modes.
In the linear stage of the instability, the sine wave corresponding to the most unstable mode.
As the amplitude increases, the sine wave is distorted by nonlinearity, and deforms into a saw tooth wave. 
The vorticity increases on the edges of the saw tooth waves and creates singular peaks.
Subsequently, each singular peak is released into each bulk, emitting a singly quantized vortex.
The release of vortices reduces the vorticity of the vortex sheet and therefore 
reduces the relative velocity across the interface. 
The released vortices drift along the interface and the system never recovers its initial flat interface.
These nonlinear dynamics differ considerably from those in classical KHI,
where the interface wave just grows into roll-up patterns.

Other kinds of instability such as Rayleigh--Taylor instability \cite{Sasaki,Gautam} and Richtmyer--Meshkov instability \cite{Bezett} are studied theoretically and numerically. 

Spinor BECs can be an important system for hydrodynamics too \cite{Kurnreview}.
Fujimoto and Tsubota investigated theoretically and numerically the GP model of spin-1 spinor BECs.
They considered the two cases: one is the counterflow of two components with different magnetic quantum numbers in a uniform system\cite{Fujimoto2012a} and the other is starting from a helical spin structure in a trapped system \cite{Fujimoto2012b} .
When the interaction is ferromagnetic, the instability is amplified to spin turbulence in both cases, where the spectrum of the spin-dependent interaction energy exhibits a -7/3 power law.
This power law is understood from some scaling argument for the equation of motion of the spin density vector. 
Since such spin density vector can be observed \cite{Vengalattore08}, such spin turbulence could be realized and observed.

\section{Conclusions}
 We review briefly the recent progresses in hydrodynamic instability and turbulence in quantum fluids.
This research started from the field of superfluid $^4$He.
Recently, important innovation occurred in the fields of both superfluid helium and atomic BECs. 
In the field of superfluid helium, scientists start to consider the classical analogue of QT. New visualization techniques are very powerful . 
In the field of atomic BECs, many kinds of hydrodynamic instability are considered theoretically and numerically, and even QT in a trapped BEC is eventually realized.
However, this article skips too many interesting and important topics because of the length constraint.
Hydrodynamics and QT in superfluid $^3$He \cite{EltsovPLTP,FisherPLTP} are not described here.
The readers should refer to other review articles \cite{Hal,TsubotaJPSJ,Vinen10,Skrbek2012,TK12}.

One of the important future works in both superfluid helium and atomic BECs would be how to characterize QT and the transition to QT. 
Then we should deal with statistical quantities.
Although energy spectra are described in this article, there are many other quantities.
For example, the probability density function of the superflow velocity is shown to exhibit some characteristic behavior different from classical case \cite{White2009,Adachi2011}.  
How QT decays is also an important problem related with quantum or quasiclassical types of QT \cite{Vinen10,Walmsley2008,Baggaley2012}.
The behavior of these quantities should be reduced to physics of quantized vortices.


\end{document}